# Testing Differences Statistically with the Leiden Ranking

**The Leiden Ranking 2011/2012 provides the *Proportion top-10% publications* ($PP_{top\ 10\%}$) as a new indicator. This indicator allows for testing the difference between two ranks for statistical significance.**

On December 1, 2011, the Centre for Science and Technology Studies (CWTS) at Leiden University launched the Leiden Ranking 2011/2012 at http://www.leidenranking.com/ranking.aspx. According to its authors, "The Leiden Ranking 2011/2012 measures the scientific performance of 500 major universities worldwide. Using a sophisticated set of bibliometric indicators, the ranking aims to provide highly accurate measurements of the scientific impact of universities and of universities' involvement in scientific collaboration." The CWTS website adds: "Compared with other university rankings, the Leiden Ranking offers more advanced indicators of scientific impact and collaboration and uses a much more transparent methodology."

This new indicator of impact is the *Proportion top-10% publications* ($PP_{top\ 10\%}$), which corresponds with the *Excellence Indicator* (*EI*) recently introduced in the SCImago Institutions Rankings (at http://www.scimagoir.com/pdf/sir_2011_world_report.pdf). Whereas SCImago uses *Scopus* data, the Leiden Ranking is based on the *Web-of-Science* data of Thomson Reuters. In addition to the stability intervals provided by CWTS, values for both $PP_{top\ 10\%}$ and *EI* can be tested statistically for significant differences from expectation, and the significance of differences in ratings between universities can be tested by using the so-called *z*-test for independent proportions (Bornmann *et al*., in press; Sheskin, 2011, pp. 656f.).

An Excel sheet can be downloaded from http://www.leydesdorff.net/leiden11/leiden11.xls into which the values for this indicator ($PP_{top\ 10\%}$) can be fed in order to obtain a *z*-value. The example in the download shows the results for Leiden University when compared with the University of Amsterdam (not significantly different; $p > 0.05$), and for Leiden University when compared with the expectation (the value is significantly above the expectation; $p < 0.001$). The values in the sheet can be replaced with values in the ranking for any university or any set of two universities.

## The *z*-test
The *z*-test can be used to measure the extent to which an observed proportion differs significantly from expectation, and whether the proportions for two institutions are significantly different. In general, the test statistics can be formulated as follows:

$$z = \frac{p_1 - p_2}{\sqrt{p(1-p)\left[\dfrac{1}{n_1} + \dfrac{1}{n_2}\right]}}$$

where: $n_1$ and $n_2$ are the numbers of all papers published by institutions 1 and 2 (under the column "$P$" in the Leiden Ranking); and $p_1$ and $p_2$ are the values of $PP_{top\ 10\%}$ of institutions 1 and 2. Furthermore:

$$p = \frac{t_1 + t_2}{n_1 + n_2}$$

where: $t_1$ and $t_2$ are the numbers of top-10% papers of institutions 1 and 2. These numbers are calculated (in the sheet) on the basis of "$P$" and "$PP_{top\ 10\%}$" provided by the Leiden Ranking. When testing observed versus expected values for a specific set, $n_1 = n_2$. In that case, $p_1$ is the value of the $PP_{top\ 10\%}$ and $p_2$ the expected value; for stochastic reasons the expectation is equal to 10% of $n_2$.

An absolute value of $z$ larger than 1.96 indicates the statistical significance of the difference between two ratings at the five percent level ($p < 0.05$); the critical value for a test at the one-percent level ($p < 0.01$) is 2.576. However, in a series of tests for many institutions, a significance level higher than five percent must be chosen because of the possibility of a family-wise accumulation of Type-I errors (the so-called Bonferroni correction; cf. Leydesdorff *et al.*, 2011).

In summary, tt seems fortunate to us that two major teams in our field (Granada and Leiden University) have agreed on using an indicator for the Scopus and WoS databases, respectively, that allows for statitistical testing of the significance of differences and positions. Of course, there remains the problem of interdisciplinarity/multidisciplinarity when ranking institutional units such as universities. This can be counteracted by field-normalization and perhaps by fractionation of the citations (1/NRef) in terms of the citing papers (Zhou & Leydesdorff, 2011).

Loet Leydesdorff [1] & Lutz Bornmann,[2]
December 17, 2011

---

[1] Amsterdam School of Communication Research, University of Amsterdam, Kloveniersburgwal 48, NL-1012 CX, Amsterdam, The Netherlands; loet@leydesdorff.net.
[2] Max Planck Society, Administrative Headquarters, Hofgartenstr. 8, 80539 Munich, Germany; Lutz.Bornmann@gv.mpg.de.